# Look Who's Talking Now: Implications of AV's Explanations on Driver's Trust, AV Preference, Anxiety and Mental Workload

*Accepted to Transportation Part C: Emerging Technologies*

Na Du
University of Michigan
nadu@umich.edu

Jacob Haspiel
University of Michigan
haspiel@umich.edu

Qiaoning Zhang
University of Michigan
qiaoning@umich.edu

Dawn Tilbury
University of Michigan
tilbury@umich.edu

Anuj K. Pradhan
University of Michigan
anujkp@umich.edu

X. Jessie Yang
University of Michigan
xijyang@umich.edu

Lionel P. Robert Jr.*
University of Michigan
School of Information
4388 North Quad
105 South State Street
Ann Arbor, MI
48109-1285
lprobert@umich.edu

*Corresponding Author
Declarations of interest:
none

## Abstract

Explanations given by automation are often used to promote automation adoption. However, it remains unclear whether explanations promote acceptance of automated vehicles (AVs). In this study, we conducted a within-subject experiment in a driving simulator with 32 participants, using four different conditions. The four conditions included: (1) no explanation, (2) explanation given before or (3) after the AV acted and (4) the option for the driver to approve or disapprove the AV's action after hearing the explanation. We examined four AV outcomes: trust, preference for AV, anxiety and mental workload. Results suggest that explanations provided before an AV acted were associated with higher trust in and preference for the AV, but there was no difference in anxiety and workload. These results have important implications for the adoption of AVs.

## 1. Introduction

Automated vehicles (AVs) have the potential to provide our society with more fuel-efficient driving (Chen et al. 2019; Katrakazas et al., 2015; Young & Stanton, 2004, 2007), reduce driving-related injuries and deaths (Eby et al., 2016; Robert, 2019) and reshape transportation and logistics (Alessandrini et al., 2015; Liu et al., 2019; Maurer et al., 2016; Talebpour et al., 2016). Despite this, there are serious concerns about whether individuals will choose to employ

AVs. One of the most central of these concerns is trust (Bazilinskyy et al., 2015; Verberne et al., 2012; Bansal et al., 2016; Zhang et al., 2019). One widely used definition of trust in the context of automation is provided by Lee and See (2004, p. 51), which states "Trust can be defined as the attitude that an agent will help achieve an individual's goals in a situation characterized by uncertainty and vulnerability." Research has shown that individuals are hesitant to cede complete control of driving to AVs, explaining why trust is one of the key challenges to widespread adoption of AVs (Haspiel et al., 2018; Du et al., 2018; Fraedrich et al., 2016; Ghazizadeh et al., 2012; Kaur & Rampersad, 2018; Petersen et al., 2018, 2019; Zhang et al., 2018).

Explanations have been shown to promote the use of automation in part by facilitating trust (Dzindolet et al., 2003; Herlocker et al., 2000; Madhavan et al., 2016; Manzey et al., 2012; Pu & Chen, 2006; Sarter et al., 1997; Sinha & Swearingen, 2002; Thill et al., 2014), yet it remains unclear whether or when they are likely to do the same for AVs. Explanations provided by automation are essentially reasons to justify why an action should or should not be taken. Explanations provide humans with transparency, which exposes them to the inner workings or logic used by the automated systems (Mercado et al., 2016; Seong & Bisantz, 2008). In general, the more humans understand the logic or rationale that underlies the automation, the less they worry, and the more they would be expected to trust and prefer to employ the automation (Dzindolet et al., 2003; Mercado et al., 2016). However, the impacts of AV explanations have been mixed. In some cases, AV explanations promoted trust in and preference for AVs (Forster et al. 2017; Koo et al. 2016), whereas in other cases they have not (Koo et al., 2015; Körber et al., 2018).

Given these gaps in the literature and the importance of the topic, we designed this study to examine the impact of AV explanations on several important outcomes. More specifically, this study examines trust in AV along with preference for the AV, anxiety and mental workload. This study addresses the following questions: (a) Do AV explanations promote drivers' trust in and preference for AV while decreasing anxiety and mental workload? (b) How do the timing and the degree of AV autonomy influence the effectiveness of AV explanations? Based on Uncertainty Reduction Theory (URT), we hypothesized that AV explanations are more effective at promoting drivers' trust in and preference for AVs and at reducing anxiety and mental workload when given *before* rather than *after* the AV acts. Again, based on URT, we also hypothesized that a lower degree of AV autonomy that requires user approval before the AV acts increases drivers' trust in and preference for the AV and decreases drivers' anxiety and increases mental workload.

The rest of this paper is organized as follows. Section 2 gives the background for the work and Section 3 develops the hypotheses to be tested. Section 4 describes the method. The results are presented in Section 5 and discussed in Section 6. Section 7 gives the limitations and future work, and the paper concludes in Section 8.

## 2. Background

In this section, we review several bodies of literature that informed and motivated our research. First, we provide a brief review of the literature on the role of explanations in automation, and then we present a more in-depth literature review on AV explanations. Next, we present the literature on the degree of autonomy in automation generally and AVs particularly. Finally, we present the outcomes expected to be associated with AV explanations as they pertain to trust, preference for AV, anxiety and mental workload.

## 2.1 Explanations and Automated Vehicles

Explanations can promote trust in automation (Dzindolet et al., 2003; Herlocker et al., 2000; Madhavan et al., 2016; Manzey et al., 2012; Pu & Chen, 2006; Sarter et al., 1997; Sinha & Swearingen, 2002; Thill et al., 2014). Previous studies have highlighted specific examples of the benefits of explanations. For example, Sarter et al. (1997) demonstrated that explanations help to avoid automation surprise and negative emotional reactions. Thill et al. (2014) found that participants preferred an awareness of why an autonomous navigation aid chose specific directions. Similarly, other studies have demonstrated that providing explanations for automation errors discouraged automation disuse and promoted trust in the automation (Dzindolet et al., 2003; Madhavan et al., 2006; Manzey et al., 2012). In addition, researchers have found that interfaces designed to provide explanations are effective at both building users' trust and confidence in the automation and promoting their acceptance of the automation (Herlocker et al., 2000; Pu & Chen, 2006; Sinha & Swearingen, 2002).

Surprisingly, explanations provided by AVs have not always translated into more trust in or preference for AVs. Table 1 summarizes four relevant studies that have investigated the impact of AV explanations in the context of a dynamic driving environment. Koo et al. (2015) examined the impact of AV explanations as well as the type of explanation on the positive emotional valence toward the AV, AV acceptance and driving performance. They separated AV explanation into *why* the AV acted and *how* the AV acted. They found that the why-only explanation led to the highest level of positive emotional valence. For AV acceptance, both the why-only and the why-and-how explanation conditions increased AV acceptance relative to the no-explanation condition. Interestingly, the why-and-how explanation condition produced the safest driving behavior while the how-only condition led to the most unsafe driving behaviors. All explanations were provided before the AV acted.

Table 1: Summary of the Impacts of Explanations on AV Outcomes

| Outcome | Study | | | | |
|---|---|---|---|---|---|
| | Koo et al., 2015 | | Koo et al., 2016 | Forster et al., 2017 | Körber et al., 2018 |
| Positive Emotional Valence | (+) Why only (Before) | (-) Why and How (Before) | | | |
| Driving Performance | (+) Why and How (Before) | (-) How only (Before) | | | |
| AV Preference /Acceptance | (+) Why only, Why and how (Before) | (-) How only (Before) | (+) Why only (Before) | | (NS) Why only (After) |
| Sense of Control | | | (+) Why only (Before) | | |
| Anxiety | | | (-) Why only (Before) | | |
| Alertness | | | (+) Why only (Before) | | |
| AV Trust | | | | (+) Why and How (Before) | (NS) Why only (After) |

| AV Usability | | | | (+) Why and How (Before) | |
| Anthropomorphism | | | | (+) Why and How (Before) | |

Note: "Before" and "After" had different explanation timing; "Why only," "How only" and "Why and How" indicate different explanation types; "+" and "-" and "NS" show the effects of explanations on outcomes compared to no-explanation conditions

Koo et al. (2016) examined in part how explanations in the form of alerts provided by the AV before action was taken impacted drivers' anxiety, sense of control, alertness and preference for AV. They found that when the AV explained what it was going to do before it acted, it decreased drivers' anxiety and increased their sense of control, alertness and preference for the AV. Similarly, Forster et al. (2017) found that an interface with speech output explaining the action the AV was going to take was rated as superior for its trust, anthropomorphism and usability when compared to interfaces that did not explain AV actions. Contrary to these positive outcomes, Körber et al. (2018) found that explanation provided after an AV requested a takeover did not significantly increase trust in the AV, although it did increase perceived understanding of the system and the reasons for the takeovers.

As summarized in Table 1, researchers have focused on different explanation types and timing in different studies. While the effects of explanation type have been compared directly in Koo et al.'s study, no study has systematically explored the effects of explanation timing.

**2.2 Degree of Autonomy**

Degree of autonomy — the degree to which the automation can make decisions and take actions independently of the user — is an important determinant of whether someone is willing to employ automation. Although there are many ways to classify the degree of autonomy, Sheridan and Verplanck (1978) put forth one of the earliest and most popular classifications with ten levels. The ten levels were based in part on whether the automation recommended a course of action and allowed the user to decide, or the automation made the decision and took the action without consulting the user. For example, at levels 4, 5 and 6, the automation suggests a course of action but only executes the suggestion if the user approves. As the automation level increases, the automation acts automatically without the user's permission. Parasuraman, Sheridan and Wickens (2008) acknowledged the Sheridan–Verplanck 10-point scale and introduced the idea of associating levels of automation to functions, where decision and action were important functions that the users and the automation shared.

Previous studies have verified the importance of the degree of autonomy by demonstrating its link to outcomes like trust in automation (de Visser & Parasuraman, 2011; Rovira et al., 2007; Verberne et al., 2012; Willems & Heiney, 2002). For example, Rovira et al. (2007) investigated the effects of imperfect automation on decision-making in a simulated command-and-control task and found that trust was higher for automation that had a lower degree of autonomy. One reason given by Verberne et al. (2012) is that users tend to trust lower rather than higher levels because they feel out of the loop as the degree of autonomy increases. This is consistent with the results of de Visser and Parasuraman (2011), which showed that automation that adapts its degree of autonomy to match the needs of a given situation and user preference leads to more trust in the automation.

Acknowledging the importance of the degree of autonomy for automated vehicles, SAE International identified six levels of driving automation (SAE International, 2018). Despite this, it should be noted that our study only considers what SAE would define as automated level 4 vehicles and above, where the AV handles all aspects of the dynamic driving task and there is no need for drivers to take control of the vehicle.

In sum, previous work has emphasized that the degree of autonomy is vital to understanding when users might trust or employ automation. However, the literature offers little insight into whether or how the degree of autonomy might influence the impact of AV explanations. Yet, based on the prior literature, we might expect the same explanation to be received differently based on whether the user can approve or disapprove the proposed course of action. Therefore, it remains vital to verify whether the degree of autonomy is indeed important to understanding the effectiveness of AV explanations. In this work, we examine two different degrees of autonomy: either the AV takes actions automatically, or it asks the participant for approval before taking an action.

## 2.3 Automated Vehicle Explanation Outcomes

Prior studies examining the impact of AV explanations have identified several important outcomes. In all, these outcomes represent either barriers to adoption or positive attitudes associated with successful adoption. These barriers or positive attitudes associated with adoption outcomes included various measures of trust, preference for AV, and anxiety. Building on prior research, we examined the influence of AV explanations on all of these outcomes as well as mental workload. Next, we present and discuss each outcome and the justification for its inclusion in our study.

Trust in automation has been shown to be vital to understanding technology use in general and AV use specifically. Trust continues to be important as the degree of complexity and uncertainty associated with new automation increase (de Vries et al., 2003; Parasuraman & Riley, 1997; Parasuraman et al., 2008; McGuirl & Sarter, 2006). Trust in automation is positively associated with both intentions to use and effective use of automation (Lee & Moray, 1992, 1994; Muir, 1987; Muir & Moray, 1996). However, problems can arise when trust is not aligned (i.e. too much or too little) with the automation's capability (Lee & Moray, 1994). Researchers have demonstrated that providing users with accurate information about the automation can assist in the development of an appropriate level of trust (Dzindolet et al., 2003; Seppelt & Lee, 2007).

Trust is particularly important for understanding the effective deployment of automated driving systems (Bao et al., 2012; Miller & Ju, 2014; Seppelt & Lee, 2007; Verberne et al., 2012). Researchers have examined various factors that influence drivers' trust in AVs. For example, Verberne et al. (2012) and Seppelt and Lee (2007) found that users had higher trust in adaptive cruise control systems that shared the driving goals of the user and provided the user with behavior information. Beller et al. (2013) and Helldin et al. (2013) showed that the presentation of AV's uncertainty information to drivers led to more trust and acceptance, better situational awareness, and better knowledge of the AV's limitations.

Many of the studies that have focused on trust in AVs also examined preference for AV and anxiety (Abraham et al., 2017; Koo et al., 2015, 2016; Molnar et al., 2018; Nass et al., 2005, Takayama and Nass, 2008; Shabanpour et al., 2018). Anxiety is often defined as a feeling of fear, worry, apprehension or concern. The more effective AV explanations are, the less anxiety someone should have about using the AV (Koo et al. 2016). Thus, anxiety should be an outcome used to assess the effectiveness of AV explanations.

Preference for AV is another important outcome. In this paper, preference is the degree to which someone likes or has a fondness for a particular AV technology. Preference is important because all things being equal, individuals may prefer one AV technology over others for reasons that are not always understood (Abraham et al., 2017). If this were true, it would be important to capture preference along with the measures of trust and anxiety. Therefore, in this study we included a measure of preference. We employed measures of anxiety and preference originally developed by Nass et al. (2005, 2008) and later adapted by Koo et al. (2015, 2016) to understand driver responses to AV explanations. Similarly, others like Abraham et al. (2017) have examined preference when studying automated driving technologies.

As mentioned, we also included mental workload as an outcome measure. Although previous studies examining AV explanations have not examined mental workload, it has been a focal point for many other AV studies (Molnar et al., 2018; Jamson et al., 2013; Young et al., 2004, 2007). For example, Molnar et al. (2018) measured drivers' mental workload during the transfer of control between automated and manual driving in a simulator study. Mental workload is likely to be particularly important because AV explanations might influence users' mental workload during automated driving, which might in turn influence ease of use (Naujoks et al., 2016). As such, we included mental workload as an outcome measure.

In all, knowing drivers' trust, preferences, anxiety and mental workload in relation to different approaches to deploying AV explanations would help designers and policymakers facilitate the adoption of AVs. These four measures are typically used to understand technology adoption in general. Other scholars may compare and contrast our results with the broader technology adoption literature. These four outcomes also allow us to directly compare and contrast our findings with the existing literature on AV explanations, and clearly articulate our study's contributions.

## 3. Hypotheses Development

Uncertainty Reduction Theory (URT) asserts that individuals seek to reduce uncertainty through information (Baxter & Montgomery, 1996). Uncertainty can be defined as the inability to determine the actions of another (Baxter & Montgomery, 1996; Kramer, 1999). As uncertainty regarding a person increases, the trust one has in that person decreases and vice versa (Colquitt et al., 2012; Robert et al., 2009). According to URT, uncertainty is decreased through the acquisition of information about that person through communication (Gudykunst & Nishida, 1984). Although URT was originally developed to explain encounters between strangers, it has also been used to understand interactions between AV and pedestrians (Jayaraman et al., 2018).

Based on URT, we derived a set of three hypotheses to help understand the effects of explanations and their timing, and the effects of degree of autonomy on drivers' trust, preference, anxiety and mental workload. First, when an AV provides an explanation for its behavior, regardless of the explanation's timing, trust in and preference for the AV should increase, while anxiety and mental workload should decrease. Simply put, explanations reduce uncertainty by providing information about the AV's behavior to the driver. Decreases in uncertainty should lead to increases in trust and are likely to be preferred by drivers. Reductions in uncertainty also decrease the concerns and effort the driver spends attempting to understand the AV's behavior.

At the same time, explanations given before the AV has acted should be associated with higher trust in and preference for the AV, with lower anxiety and mental workload when compared to explanations given after the AV has acted. *Before* explanations are likely to head off any concerns the driver has about the actions of the AV. *Before* explanations not only provide information about the AV's actions, but also give the driver time to prepare and expect the AV's actions to take place. This reduces the chance of the AV's actions startling and possibly even alarming the driver.

Finally, we expect that allowing the driver to disapprove the AV's actions will lead to the highest levels of trust, preference and mental workload while leading to the lowest level of anxiety. Allowing the driver a restricted veto time before automatic execution should result in the highest reduction in uncertainty regarding the AV's action. However, we do expect mental workload to increase because the driver would be forced to make decisions regarding the AV's actions, which requires more attention and effort on the part of the driver. Based on these arguments, we derived a set of hypotheses.

> *H1: AVs that provide explanations have higher driver (a) trust and (b) preference along with lower driver (c) anxiety and (d) mental workload than AVs that do not provide explanations.*

> *H2: AVs that provide explanations before acting have higher driver (a) trust and (b) preference along with lower driver (c) anxiety and (d) mental workload than AVs that provide explanations after acting.*

> *H3: AVs with a lower degree of autonomy that gives the driver an option to disapprove the AV's actions have higher driver (a) trust, (b) preference and (c) mental workload along with lower driver (c) anxiety than AVs that execute the action without asking permission.*

## 3. Method

To test the hypotheses, we conducted an experiment in a controlled lab setting using a high-fidelity driving simulator. We manipulated explanation timing (before vs. after) and degree of AV autonomy (driver's permission needed vs. no permission needed) and examined their impacts on all four outcome variables: trust, preference for AV, anxiety and mental workload. This section provides details about our study.

## 3.1 Participants

Thirty-two people (11 females) with an average age of 26.9 years (SD = 6.3 years) participated in the experiment. We screened them for various inclusion criteria including driver's license status and susceptibility to simulator sickness. Participants were paid $20 for their participation in the 60- to 75-minute study. This research complied with the American Psychological Association Code of Ethics and was approved by the institutional review board at the University of Michigan. We obtained informed consent from each participant.

## 3.2 Apparatus

We conducted this study in a high-fidelity advanced driving simulator at the University of Michigan Transportation Research Institute (UMTRI) (see Figure 1). UMTRI's fixed-base simulator consists of a Nissan Versa sedan located in a dedicated lab space, integrated with a simulation system running version 2.63 of Realtime Technology's (RTI) simulation engine SimCreator along with custom coding for automated vehicle features. To present the virtual driving environment to participants, forward road scenes are projected onto three screens about 16 feet in front of the driver (120-degree field of view) and a rear screen 12 feet away (40-degree field of view). Each forward screen is at a resolution of 1,400 x 1,050 pixels and updates at 60 Hz, and the rear screen is set at a resolution of 1,024 x 768 pixels. For this study, the automation features of the driving simulator were programmed to simulate an SAE level 4 AV, wherein the longitudinal and lateral vehicle control, navigation, responses to traffic control devices and other traffic elements were all undertaken by the AV, with the driver not required to actively monitor the environment (SAE International, 2018). The simulation could be driven in non-automation mode, but for the purposes of this study, after starting a simulated drive, the participant was instructed to engage automation (via a button on the steering wheel), after which he or she was never asked to take back control of the drive.

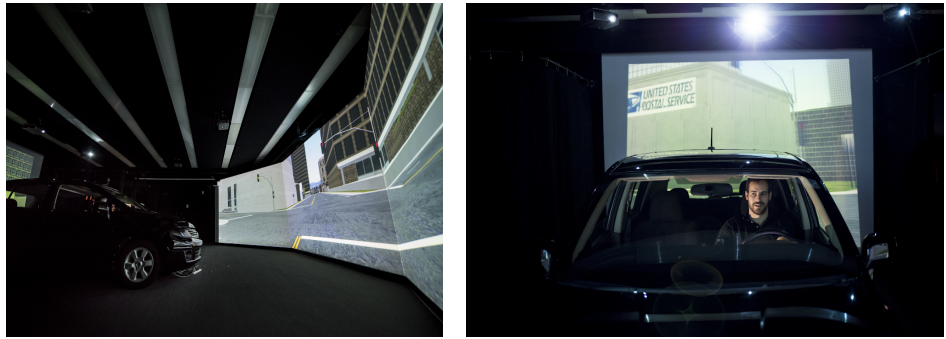

Figure 1. Driving simulator

### 3.3 Experimental Design

This study employed a within-subjects experimental design with four AV explanation conditions. The first condition was no explanation (NExpl), with the AV providing no explanation about its actions. The second condition was before explanation (BExpl), with explanations presented 7 seconds prior to the AV's actions. The third condition was after explanation (AExpl), with explanations presented within 1 second after actions had been taken by the AV. In the fourth condition (PermReq), the AV providing an explanation for its upcoming action and then, 7 seconds later, asked the driver to approve or disapprove the action. This condition examined the impact of the degree of autonomy by lowering the AV's ability to take action independently of the driver's approval. If the driver disapproved, the AV did not engage in the action. The permissions for the AV's actions were delivered via participants' verbal input. All the explanations had the same structure and wording with the only difference being the respective cause and action (see Table 2). The explanations were presented with a neutral tone of male voice with a standard American inflection. The within-subjects experimental design controlled for the individual differences, and the sequence of the four AV explanation conditions was counterbalanced using a Latin square design.

In each AV explanation condition, participants engaged in a 6- to 8-minute drive without the need to take over control of the vehicle. None of the participants in the study took over control of the vehicles, as instructed. As shown in Table 2 and Appendix 1, each drive contained three unexpected events in the environments of urban, highway and rural: events by other drivers, events by police vehicles, and unexpected re-routes. The events were chosen from previous literature and corresponded to realistic unexpected situations in automated driving (Koo et al., 2016; Lenné et al., 2008; Merat & Jamson, 2009; Miller et al., 2014; Molnar et al., 2018; Rezvani et al., 2016). All the events were programmable considering the accessibility of the driving simulator. The AV would take unexpected but reasonable actions in every event. Events occurred at prescribed times in the drive with an interval of 1–2 minutes. Each event was distinct regarding the surrounding environment, and the order of event type was counterbalanced via a Latin square design across four AV explanation conditions.

## 3.4 Dependent Variables

The dependent variables in this study are participants' subjective attitudes. The attitudinal measures include trust, mental workload, anxiety and preference. The preference and anxiety questionnaire was adapted from a published model from the CHIMe Lab at Stanford University that is used to measure driver attitude (Koo et al., 2016; Nass et al., 2005; Takayama & Nass, 2008). Anxiety comprised the averaged responses to four adjective items to describe the AV experience: fearful, afraid, anxious and uneasy. Preference for AV comprised the averaged responses to eight items: intelligent, effective, reliable, helpful, smart, dependable, high quality and efficient. All the items were rated on seven-point rating scales (1: describes very poorly; 7: describes very well). Meanwhile, trust was measured using 7-point Likert scales with six dimensions (Muir, 1987): competence, predictability, dependability, responsibility, reliability and faith (1: not at all; 7: extremely well). The Muir questionnaire represents a highly validated trust-in-automation scale and is comparable with the Jian trust scale (Desai et al., 2012). We adapted the scale to reflect the driver–AV interaction context. Participants also ranked each AV explanation condition on trust from 1 (most trust) to 4 (least trust). Finally, as a measure of mental workload, we used a subjective mental workload assessment tool, NASA-TLX, with a scale from 0 to 20 (Hart & Staveland, 1988). NASA-TLX was developed with six subscales to represent mental, physical, and temporal demand; frustration, effort and performance. We used a modified version where the subscales were averaged without the paired comparisons (Hart, 2006). All the questionnaires we employed in the study are included as appendices.

Table 2. Event descriptions.

| AV Explanation Conditions | Events | Explanations |
|---|---|---|
| No explanation (NExpl) | Efficiency Route Change | No explanations |
| | Swerving Vehicle Ahead | |

| | Stopped Police Vehicle on Shoulder | |
|---|---|---|
| Before explanation (BExpl) | Oversized Vehicle Ahead | "Oversized vehicle blocking roadway. Slowing down." |
| | Heavy Traffic Rerouting | "Rerouting, traffic reported ahead." |
| | Police Vehicle Approaching | "Emergency vehicle approaching. Stopping." |
| After explanation (AExpl) | Stopped Police Vehicle on Shoulder | "Emergency vehicle on shoulder, changed lanes." |
| | Abrupt Stopped Truck Ahead | "Roadway obstruction, changed lanes." |
| | Road Hazard Rerouting | "Rerouted. Identified road obstruction." |
| Permission required (PermReq) | Police Vehicle Approaching | "Emergency vehicle approaching, pull over and stop?" |
| | Unclear Lane Markings Rerouting | "Unclear lane lines, reroute?" |
| | Vehicle with Flashing Hazard Lights Ahead | "Vehicle with hazard lights ahead, slow down?" |

## 3.5 Procedure

Participants came to the advanced driving simulator lab and filled out the consent form acknowledging understanding of the demands of participating in the study. Upon giving consent, participants completed a web-based demographics questionnaire on an iPad. After debriefing (see Appendix 2), we conducted an initial training session to help the participants get familiar with the driving environment before the main study started. Following instructions, we prompted participants to drive as they normally did and showed them how to transfer the AV from manual control to automated mode. Participants also practiced giving permission to AV's actions via their verbal input. Then the AV would respond accordingly.

After the 3-min training portion of the study, participants experienced approximately 60 minutes in four AV explanation conditions. Each drive consisted of three events and lasted 6–8 minutes. Each drive was differentiated by explanation timing and degree of autonomy, as discussed. The order of drive for each participant was determined by the Latin square design to avoid any order effect. Participants had 1–2 minutes of rest after each drive. At the end of each drive, we asked participants to fill out a survey on the iPad to measure their trust, preference, anxiety, and mental workload, as described, based on their experience in the experimental condition (Appendices 3, 4, 5 and 6). After all drives had been completed, a trust ranking was recorded to examine participants' relative trust in the four different AV explanation conditions (Appendix 7).

## 4. Results

### 4.1 Measurement validity

To assess discriminant and convergent validity, we used the square root of the average variance extracted (AVE) values (see Table 3). To assess convergent validity of a construct the square root of the AVE should be higher than .50 (Fornell & Larcker, 1981). When the AVE value is above .50, the variance explained by the construct is greater than the variance explained by measurement error, which indicates evidence of convergent validity of the construct. The AVEs of trust ratings, preference, mental workload and anxiety were .80 and .60, .74, and .79, respectively, all of which were above .50 as recommended by Fornell and Larcker (1981). To assess discriminant validity, we compared the correlations of all constructs with the square root of the AVE values of trust ratings, preference, mental workload and anxiety. The correlation matrix, shown in Table 3, indicates that except preference, correlations among constructs were well below the square root of AVEs, which is further evidence of discriminant validity among dependent measures. Additionally, reliabilities of trust ($\alpha = 0.923$), preference ($\alpha = 0.885$), anxiety ($\alpha = 0.893$) and workload ($\alpha = 0.754$) exceeded the 0.7 recommendation (Carmines & Zeller, 1979; Fornell & Larcker, 1981).

Table 3. Correlation Matrix.

|  | Mean | Std. Deviation | Trust Rating | Trust Ranking | Preference | Mental workload | Anxiety |
|---|---|---|---|---|---|---|---|
| Trust Rating | 5.64 | 0.96 | (.80) | -0.20 * | 0.78 *** | -0.23 ** | -0.32 *** |
| Trust Ranking | 2.50 | 1.12 |  | NA | -0.24 ** | 0.07 | 0.11 |
| Preference | 5.44 | 0.92 |  |  | (.60) | -0.23 ** | -0.39 *** |
| Mental workload | 3.50 | 2.90 |  |  |  | (.74) | 0.20 * |
| Anxiety | 2.59 | 1.25 |  |  |  |  | (.79) |

Note: *** Correlation is significant at the 0.001 level (2-tailed); ** Correlation is significant at the 0.01 level (2-tailed); * Correlation is significant at the 0.05 level (2-tailed).

Values on the diagonals within the parentheses represent the square root of the AVE for each factor.

## 4.2 Hypothesis testing

We tested the hypotheses with data from all 32 participants. Statistical analysis was performed using IBM SPSS Statistics software. One-way repeated measures analysis of variance (ANOVA) was used because each of the 32 participants experienced all four of the AV explanation conditions. The statistical analyses examined the relationship between independent variables (four AV explanation conditions) and dependent variables (subjective attitudes in the AV including trust, preference, anxiety and mental workload). Note that there were two measures of trust: the survey (Appendix 3), and the ranking. Since the trust ranking of the four drives was ordinal, we used the non-parametric method, the Friedman test, to examine the differences of trust ranking among drives. The alpha level was set at .05 for all statistical tests. All post hoc comparisons utilized a Bonferroni alpha correction.

### 4.2.1 The effects of explanation timing and degree of autonomy on trust ratings and rankings

As shown in Figure 2, there was a main effect of AV explanation conditions on trust ratings in AVs ($F(3,93) = 4.814$, $p = .008$). Post hoc comparisons revealed that participants tended to have the highest trust in the AV when the explanation was given before the action (BExpl vs. NExpl: $p = .013$; BExpl vs. AExpl: $p = .018$; BExpl vs. PermReq: $p = .04$). In order to decide whether providing explanations promoted higher trust regardless of timing, we averaged the trust ratings of BExpl and AExpl and compared them to NExpl. The results showed that there was no significant difference between no-explanation and explanation-provided conditions ($F(1,31)$ = .506, $p = .482$) on trust ratings.

In addition, the Friedman test showed that the trust rankings of four AV explanation conditions were significantly different ($\chi2(3)= 49.462$, $p < .001$). To be more specific, post hoc comparisons indicated that participants gave a higher trust ranking when they were presented with the BExpl and PermReq than NExpl and AExpl (BExpl vs. NExpl: $p < .001$; BExpl vs. AExpl: $p < .001$; PermReq vs. NExpl: $p < .001$; PermReq vs. AExpl: $p < .001$). Nonetheless, there was no significant difference of trust ranking between PermReq and BExpl ($Z = -1.020$, $p = .308$).

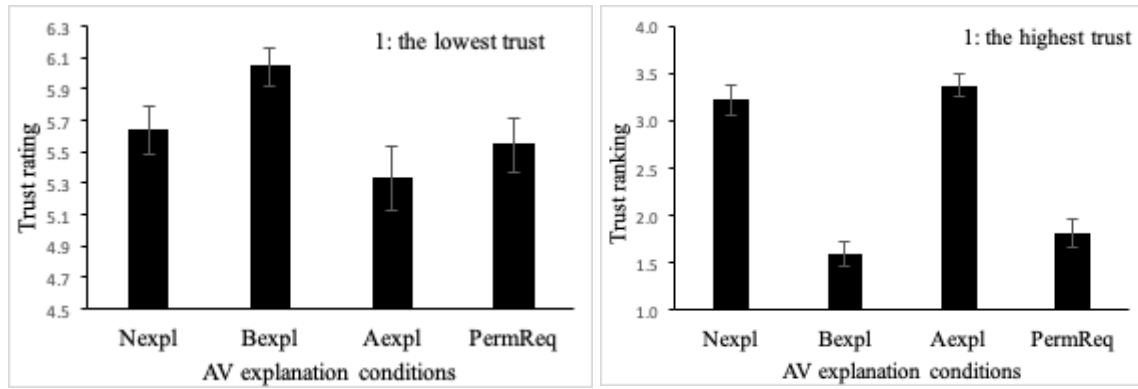

Figure 2. The average of trust rating (1-7) and ranking (1st-4th)

### 4.2.2 The effects of explanation timing and degree of autonomy on preference

The main effect of AV explanation conditions on driver preference was significant ($F(3,93) = 3.661$, $p = .027$). As illustrated in Figure 3, post hoc analysis indicated that participants preferred BExpl to NExpl ($p = .013$) and to PermReq ($p = .031$). In order to decide whether providing explanations engendered higher preference independent of timing, we averaged the preference ratings of BExpl and AExpl and compared them to NExpl. The results showed that there were no significant differences between no-explanation and explanation-provided conditions ($F(1,31) = 1.279$, $p = .267$) on driver preference.

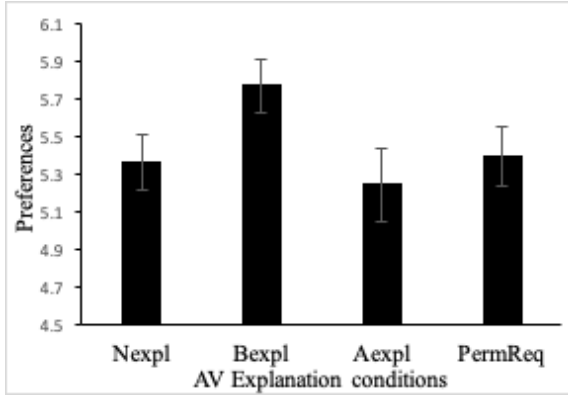 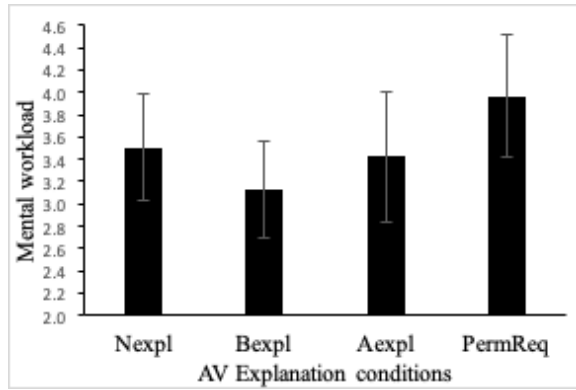

Figure 3. The average scores of preference    Figure 4. The average scores of mental workload

### 4.2.3 The effects of explanation timing and degree of autonomy on mental workload

The average scores of mental workload survey items were not significantly different among four AV explanation conditions ($F(3,93) = 2.233$, $p = .09$). However, the mental workload in BExpl had the lowest mean, as illustrated in Figure 4.

### 4.2.4 The effects of explanation timing and degree of autonomy on anxiety

As shown in Figure 5, no significant differences among four AV explanation conditions were found in driver's anxiety toward the AV ($F(3,93) = .525$, $p = .666$).

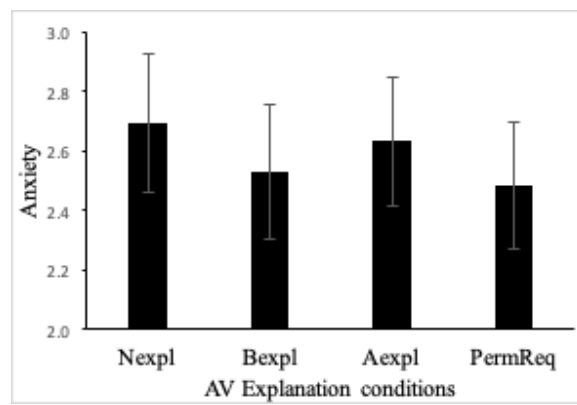

Figure 5. The average scores of anxiety

## 4.3 Summary of the Results

The findings from this paper can be organized into three overarching results. One, this study found no evidence that AVs that provided explanations led to higher trust and preference along with lower anxiety and mental workload than AVs that did not provide explanations. Thus, H1 was not supported. Two, this study did find some significant outcome differences between providing explanations before and after the AV has taken actions. Thus, H2a and H2b were supported, while H2c and H2d were not. Three, results of this study found no evidence that AVs with a lower degree of autonomy that provided the driver an option to disapprove their actions increased preference and mental workload or lowered anxiety compared to AVs that did not give drivers options. However, while trust ratings in the AV were not significantly higher when we increased the degree of autonomy, the trust rankings were higher than in the conditions that provided no explanation or provided explanations after the AV had acted. Thus, H3a was partially supported, while H3b, H3c and H3d were not. The next section provides a detailed discussion of the findings and their contributions to the literature along with the study's limitations.

Before we discuss the implications of our work, we explain here some of our non-significant findings by conducting a post hoc power analysis. Statistical power is the probability of obtaining a significant p-value with a given sample size and a given effect size (e.g., differences in means across conditions; Cohen, 1992; Ellis, 2010). According to Cohen (1988), power analysis with values of .80 or above are considered powerful enough to detect medium effect sizes. In this study, our significant p-value was less than 0.05 ($p < .05$). Originally, we anticipated a medium effect size for our within-subject design. To calculate the sample size for a repeated measures design, we used G*Power (see Faul et al., 2007, 2009). Results indicated that a sample size of 32 would provide a power of .92, well above the .80 suggested by Cohen. Therefore, for a medium effect size, our sample size of 32 for a repeated measure design provided sufficient power.

## 5. Discussion

This study contributes to the literature in the following ways. First, it demonstrates that timing matters when understanding the impacts of AV explanations. The overall mean trust of both explanation conditions was not significantly different from the no-explanation condition. Apparently, just providing an explanation is not sufficient for increasing trust. Our results help clarify why several studies found that AV explanations did increase trust and one study did not. The studies that found support for explanations increasing trust had the AVs provide them before acting (see Forster et al., 2017; Koo et al., 2015, 2016), whereas the study where the AV provided an explanation after it had already acted (see Körber et al., 2018) did not demonstrate increased trust. By directly comparing the effects of the no-explanation with the before- and after-explanation conditions, this study can draw the conclusion that merely providing an explanation is not enough; the AV needs to provide the explanation before acting.

It is important to note that the AV simulated in our study can be considered as level 4 automation. Level 4 automation is where the AV is able to "perform all safety-critical driving functions and monitor roadway conditions for an entire trip" (National Highway Traffic Safety Administration [NHTSA], 2013, p. 5). However, the AV automation level of the prior literature simulated level 2 and level 3 automation (Forster et al., 2017; Koo et al., 2015, 2016; Körber et al., 2018). Nonetheless, our results are consistent with prior studies simulating level 2 and level 3 automation, that providing explanations before instead of after AV action promotes trust.

With regard to preference, the results of our study are also consistent with studies of AVs that provide explanations before they act. Our findings align with those of Koo et al. (2015, 2016) and Forster et al. (2017), who found that providing *before* explanations promoted preference for the AV. It would appear that individuals prefer AVs that provide explanations before acting. Similar to our findings on trust in the AV, no such effects were associated with AVs that provided the explanation after taking action. The results are also consistent with the high correlation between trust and preference (see Table 3).

However, our findings also differ from those of Koo et al. (2016) with regard to decreases in anxiety. In contrast to Koo et al., our study shows that providing explanations did not significantly affect drivers' anxiety. Anxiety was not significantly lower for either the before- or after-explanation condition when compared to the no-explanation condition. Our differences with regard to anxiety might have been caused by the different automation levels used in the two studies. In the study of Koo et al. (2016), the driver was involved in the low-level control of the vehicle and an emergency braking system was activated in impending collision situations. In such a setting, drivers could easily perceive the discrepancy between the low-level control and the vehicle's behavior, resulting in higher anxiety if an explanation was not provided. On the contrary, in our study, the higher level of autonomy (level 4) did not require the driver to perform any driving actions. This might have reduced any driver concerns or anxiety.

Second, this study found little evidence that increasing the user's control by lowering the degree of autonomy mattered. The lower degree of autonomy condition, which asked for the driver's permission to act, did not lead to an increase in trust, preference or mental workload, nor did it reduce anxiety. The are several possible reasons for this. One, the driver might have thought that the AV that asked for permission was less capable because it could not determine what to do on its own. Our preference measure had items asking participants how intelligent they thought the AV was for each condition. We found that participants in this experiment viewed the AV asking for permission as less intelligent when compared to the AV that provided the before-action explanation ($p = .041$). Two, participants might have found the need to continuously judge the situation and make decisions within a limited time too mentally taxing or burdensome. This might be explained by Bainbridge (1983), who pointed out the ironies of automation in that operators sometimes find it difficult to handle the monitoring and coping of unusual circumstances that are required with greater control over the automation. Although this was not significant at the .05 level, participants did report higher levels of mental workload in the permission condition.

The results were less clear with regard to trust by rankings. The condition that provided greater control through lower degree of autonomy ranked higher than the no-explanation and after-explanation conditions. The differences in our measure of trust via attitude versus the measure of trust via ranking might explain this finding. Trust measured via attitude is defined by items we obtained from the literature (see Muir, 1987). These items specifically mention competence, predictability, dependability, responsibility, reliability and faith. However, trust measured via ranking might represent how participants define trust. The participants' idea of trust might not have aligned with our measure of trust in this study. This potential disconnect between the measures might warrant further investigation into when and why these measures are likely to diverge.

Finally, this study highlights the benefits and limitations associated with employing URT to understand both AV explanations and degree of autonomy on trust, preference, anxiety and mental workload. URT did not explain why explanations in general did not have an overall significant effect on our outcomes. Ideally, providing explanations should demonstrate benefits over not providing explanations. The more information gained through communication, the less uncertainty one has about the AV. Explicitly explaining the *why* should have reduced the uncertainty and increased the transparency of the AV's actions, leading to higher trust and preference. But our results do not support this assertion.

URT does explain why before-explanations were significantly better than no-explanation and after-explanations. Explicitly explaining why the AV was acting before it acted reduced the uncertainty and increased the transparency of the AV's actions, leading to higher trust and preference for the AV, and lower mental workload and anxiety. These findings are all supported by URT. Finally, URT did not explain why more control over the AV's actions did not lead to significant benefits, with the exception of trust rankings. We expected that the lower degree of autonomy that gave the driver an option to disapprove the AV's actions would lead to the lowest level of uncertainty, and therefore the highest trust, preference and mental workload with the lowest anxiety. This was not supported. In all, our results provide mixed support for the potential of URT to help us understand the impacts of AV explanations and degree of autonomy.

## 6. Limitations and Future Work

Findings of the present study should be interpreted in light of the following limitations. First, even though the study was conducted in a high-fidelity driving simulator, it still reduced the risk of unexpected events in highly automated driving, which might influence participants' feelings such as anxiety and perceived safety. Second, the preference survey we employed in the study was highly correlated with trust rating and didn't have a high discriminant validity. It might result from similar meanings between trust and preference. Individuals prefer things they trust and trust things they prefer. Third, the results we obtained are general findings on average of all participants. We did not consider individual differences such as desirability of control and personality. Future study could propose an adaptive interface whose explanation timing and degree of autonomy can be adjusted in response to drivers' characteristics. Fourth, explanations we used in the study were all auditory. Future studies might examine multimodal explanations. Finally, we did not collect qualitative measures of our various outcomes. Future studies should consider including qualitative measures to provide additional insights.

## 7. Conclusion

This study investigated the effects of the timing of AV explanations and the degree of autonomy on drivers' trust, preference for AV, anxiety and mental workload. Our findings extend previous work in the following ways. First, we focused on SAE level 4, which means the driver is no longer required to keep his or her hands/feet on the steering wheel/pedals and is permitted to take eyes off the road for extended periods during highly automated driving (SAE International, 2018). Second, we identified and demonstrated the importance of the timing of AV's explanations. Finally, we went beyond existing literature on AV explanations by incorporating the impacts of the degree of autonomy. This study is an important contribution to the literature on AVs. Nonetheless, more research is needed to build and expand on these ideas to further provide new insights.

## Acknowledgment

This work was supported by University of Michigan Mcity and in part by the National Science Foundation.

**Appendix 1**

| Events | Descriptions |
|---|---|
| Efficiency Route Change | The AV rerouted in view of road construction ahead. |
| Swerving Vehicle Ahead | The vehicle ahead was swerving, so the AV slowed down until the swerving vehicle exited the highway. |
| Stopped Police Vehicle on Shoulder | A police vehicle stopped on shoulder, so the AV changed lane to avoid collision. |
| Oversized Vehicle Ahead | There was an oversized vehicle ahead blocking roadway, so the AV slowed down until the oversized vehicle took turns at the intersection. |
| Heavy Traffic Rerouting | Heavy traffic jam was reported ahead, so the AV rerouted. |
| Police Vehicle Approaching | A police vehicle approached the AV from behind and activated siren. Then the AV pulled over and stopped. |
| Stopped Police Vehicle on Shoulder | A police vehicle stopped on shoulder, so the AV changed lane to avoid collision. |
| Abrupt Stopped Truck Ahead | There was roadway obstruction ahead. Then the AV changed lanes. |
| Road Hazard Rerouting | The AV rerouted because it identified road hazard ahead. |

| Police Vehicle Approaching | A police vehicle approached the AV from behind and activated siren. Then the AV asked driver's permission whether to pull over. |
|---|---|
| Unclear Lane Markings Rerouting | When the AV approached the intersection, the lane marking ahead was not clear. Then the AV asked driver's permission whether to reroute. |
| Vehicle with Flashing Hazard Lights Ahead | A vehicle in the left front lane was flashing the hazard light. Then the AV asked driver's permission whether to slow down. |

## Appendix 2

## Debriefing before training session

We will begin the training portion of the study. During participation in our study, you will act as the driver of a fully autonomous vehicle. This means the vehicle is able to drive safely entirely on its own. The car is able to function in all driving situations as well as the average human driver. It obeys all traffic laws. Also, it receives navigation information from external sources similar to Google Maps, and can change routes to reach a destination more quickly if one is identified or available. The autonomous vehicle maintains lanes by visually sensing the lane lines on the roadway.

For the purposes of our study once the automated driving mode has been engaged you will be unable to take control of the wheel or control the vehicle. Once automated mode is activated, you will no longer need to actively monitor the roadway or control the vehicle. Now we will begin the training simulation. It will take around 2 minutes. As in a normal car, your vehicle can be manually controlled with the steering wheel and pedals. The right pedal is your gas pedal and the left pedal is the brake. During the training session I will ask you to place the vehicle into automated mode by using the button located on the lower right part of the steering wheel labelled as ON/OFF.

You will note that the automated mode is activated because of a color indicator on the dashboard as well as an audio alert. Once you have activated automated driving mode please remove your hands from the wheel and do not place them back on the wheel. The wheel turns on its own and could potentially harm your hand if it were in the wrong position on the wheel while turning. Finally the vehicle may ask for your input in a decision via a spoken response. If prompted please respond vocally.

## Debriefing before each drive

We will now begin the main part of the study. There will be four driving sections, each followed by a survey. Each drive will be differentiated by how the autonomous vehicle presents information to you about both the environment and the actions of other vehicles on the roadway. Your goal during this part of the study is to evaluate your experience in the autonomous vehicle in each drive. Do you have any questions before beginning?

# Appendix 3

## Post-drive Trust Survey

Please indicate the extent to which you believe the autonomous vehicle has each of the following traits (from 1 representing "none at all" to 7 representing "extremely high")

1. Competence: To what extent does the autonomous vehicle perform its function properly?
2. Responsibility: To what extent does the autonomous vehicle perform the task it was designed to do? (In other words, to what extent does the autonomous vehicle drive safely?)
3. Reliability over Time: To what extent does the autonomous vehicle respond similarly when it encounters similar circumstances at different points in time?
4. What is your degree of faith that the autonomous vehicle will be able to cope with other driving states in the future?
5. Predictability: To what extent can the autonomous vehicle's behavior be predicted from moment to moment?
6. Dependability: To what extent can you count on the autonomous vehicle to do its job?
7. What is your overall degree of trust in the autonomous vehicle?

**Appendix 4**

## Post-drive Preference Survey

How well do the following adjectives describe the autonomous vehicle in the drive?

| | Extremely poorly | Very poorly | Poorly | Neutral | Well | Very well | Extremely well |
|---|---|---|---|---|---|---|---|
| Intelligent | o | o | o | o | o | o | o |
| Efficient | o | o | o | o | o | o | o |
| Smart | o | o | o | o | o | o | o |
| High Quality | o | o | o | o | o | o | o |
| Reliable | o | o | o | o | o | o | o |
| Dependable | o | o | o | o | o | o | o |
| Effective | o | o | o | o | o | o | o |
| Helpful | o | o | o | o | o | o | o |

**Appendix 5**

## Post-drive Anxiety Survey

How well do the following adjectives describe how you felt while the AV was driving itself?

|         | Extremely poorly | Very poorly | Poorly | Neutral | Well | Very well | Extremely well |
|---------|------------------|-------------|--------|---------|------|-----------|----------------|
| Anxious | o | o | o | o | o | o | o |
| Fearful | o | o | o | o | o | o | o |
| Afraid  | o | o | o | o | o | o | o |
| Uneasy  | o | o | o | o | o | o | o |

# Appendix 6

## Post-drive Mental Workload Survey

NASA TLX is a subjective mental workload assessment tool. Please mark your answers on the scales below.

Mental Demand How mentally demanding was the task?

Very Low                                                                     Very High

Physical Demand How physically demanding was the task?

Very Low                                                                     Very High

Temporal Demand How hurried or rushed was the pace of the task?

Very Low                                                                     Very High

Performance How successful were you in accomplishing what you were asked to do?

Very Low                                                                     Very High

Effort How hard did you have to work to accomplish your level of performance?

Very Low                                                                     Very High

Frustration How insecure, discouraged, irritated, stressed, and annoyed were you?

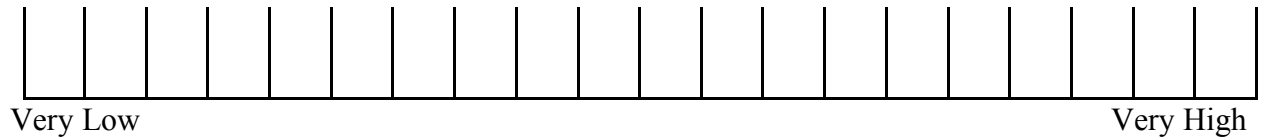

Very Low                                                                                        Very High

## Appendix 7

## Trust Ranking Survey

In the four driving sections (three videos clips in each section) you just watched, the autonomous vehicle provided information differently to you about events on the roadway. The four methods (not in order of how you experienced them) are listed below: Method A) No information provided; Method B) Information provided before events; Method C) Information provided after events; Method D) Information provided before events with driver input requested. By dragging the options listed below up or down, please rank the methods by how much you trusted the vehicle based on how it provided information to you about events on the roadway. 1 (top) corresponds to trusted the most. 4 (bottom) corresponds to trusted the least.

_______ A) No information provided

_______ B) Information provided before events

_______ C) Information provided after events

_______ D) Information provided before events with spoken input deciding vehicle action